\documentclass[11pt, reqno, letter]{amsart}
\usepackage{setspace}
\usepackage[foot]{amsaddr}
\usepackage{xcolor}
\usepackage{bm}
\usepackage{bbm}
\usepackage{graphicx}
\usepackage{tikz-qtree}
\usetikzlibrary{positioning}
\usepackage[per=slash]{siunitx}
\usepackage{natbib}
\usepackage{algorithm, algpseudocode}
\usepackage{caption}
\usepackage{multirow}

\allowdisplaybreaks

\title[Causality in the era of data science]{From controlled to undisciplined data: estimating causal effects in the era of data science using a potential outcome framework}

\begin{document}

\begin{abstract}

This paper discusses the fundamental principles of causal inference – the area of statistics that estimates the effect of specific occurrences, treatments, interventions, and exposures on a given outcome from  experimental and observational data. We explain the key assumptions required to identify causal effects, and highlight the challenges associated with the use of observational data. We emphasize that experimental thinking is crucial in causal inference. The quality of the data (not necessarily the quantity), the study design, the degree to which the assumptions are met, and the rigor of the statistical analysis  allow us to credibly infer causal effects. Although we advocate leveraging the use of big data and the application of machine learning (ML) algorithms for estimating causal effects, they are not a substitute of thoughtful study design. Concepts are illustrated via examples.

\end{abstract}

\author{Francesca Dominici, Falco J. Bargagli-Stoffi and Fabrizia Mealli}	

\keywords{causal inference, potential outcomes, study design, machine learning}

\maketitle
	
\doublespace

\section{Introduction}

Questions about cause and effect are ubiquitous in our everyday lives:
\begin{itemize}
\item Starting now, if I stop eating ice cream at night, how much weight will I lose in three months? 
\item If I adopt a puppy, will the severity of my depression symptoms improve in one year?
\item If I give my patient  new chemotherapy instead of the standard chemotherapy, how many more months will he/she live?
\item If my government implement stricter regulatory policies for air pollution, how much longer can I expect to live?
\item If my country or state had implemented a mask mandate three months ago to slow down the spread of COVID-19, how many lives would have been saved?  
\end{itemize}
These are just a few situations ranging in complexity and importance where we would like to estimate the causal effect of a defined intervention $W$ (e.g., not eating ice cream, adopting a puppy, taking a new drug, implementing air pollution regulations, enacting a stricter social distancing measure) on a specific outcome $Y$ (e.g., body weight, depression symptoms, life expectancy, COVID-19 deaths). 

This paper discusses the fundamental ideas of causal inference under a potential outcome framework \citep{neyman1923,rubin1974estimating,rubin1978bayesian} in relation to new data science developments. As statisticians, we focus on study design and estimation of causal effects of a specified, well-defined intervention $W$ on an outcome $Y$ from observational data.

The paper is divided into eight sections:
\begin{itemize}
\item Sections 1 and 2 - which heuristically contrast randomized controlled experiments with observational studies. 
\item Section 3 - the design phase of a study, including the illustration of the key assumptions required to define and identify a causal effect.
\item Section 4 - a non-technical overview of common approaches for estimating a causal effect, focusing on Bayesian methods. 
\item Section 5 - advantages and disadvantages of the most recent approaches for machine learning (ML) in causal inference.
\item Section 6 - recent methods for estimating heterogeneous causal effects. 
\item Section 7 - a discussion of the critical role of sensitivity analysis to enhance the credibility of causal conclusions from observational data. 
\item Section 8 - a concluding discussion outlining future research directions.
\end{itemize}

\subsection{The Potential Outcomes framework is just one of the many popular approaches to causal inference:} Before we present our panoramic view of the potential outcome framework for casual inference, we want to acknowledge that {\it causation} and {\it causality} are scientific areas that span many disciplines including, but not limited, to statistics. Several approaches to causality have emerged and have become popular in the areas of computer, biomedical, and social sciences.

Although the focus of this paper is on the potential outcome framework, we want to stress the importance of acknowledging the other approaches and viewpoints that exist in the general area of causality. A Google search on causality resulted in 45 books (\text{https://www.goodreads.com/shelf/show/causality}) of which less than 10 focus on statistical estimation of causal effects. Many of those books provide philosophical views of causal reasoning in the context of different disciplines and a comprehensive overview of causality as a discipline. Texts that discuss causal methods from the potential outcome perspective (even if they are not always exclusive) include \cite{morganwinship2007, morganwinship2014, angrist_mostly_2008, rosenbaum2002observational, rosenbaum2010design,imbens2015causal, hernan_robins_2020}. 
\cite{pearl2000book} and \cite{pearl2018book} approach to causality has become very popular in computer science \cite[see for example][]{Peters2017}. In particular, the books by Pearl introduce the notion of causal graphs which are graphical representations of causal models used to represent assumptions about potential causal relationships. These graphs represent assumptions regarding causal relationships as edges between nodes.  In these instances graphs are used to determine if data can identify causal effects and visually represent assumptions in causal models. The benefits of this approach are particularly evident when multiple variables could have causal relationships in a complex fashion \cite[see also][]{hernan_robins_2020}.

Other books cover the causality-philosophical meaningfulness of causation \citep{cartwright2007hunting, BeebeeHitchcockMenzies2009, halpern2016actual}; deducing the causes of a given effect \citep{dawid2017probability}; understanding the details of a causal mechanism \citep{vanderweele2015explanation}; or discovering or unveiling the causal structure \citep{Spirtesetal2001, Glymour:Scheines:Spirtes:2014}. Unfortunately, it is impossible to summarize all of these contributions in a single paper. 

For the type of studies we have encountered in our work in sociology, political science, economics, environmental, and health sciences the typical setting is to estimate the causal effect of a pre-specified treatment or intervention $W$ on an outcome $Y$. In these instances we have found the potential outcome approach useful to draw causal inferences. The potential outcome framework is also helpful to bridge experimental and observational thinking. 

In this paper we provide a statistical view of the potential outcome framework for causal inference. We emphasize that there is a lot we can learn  from the design of randomized controlled trials (RCTs) for estimating causal effects in the context of observational data. Furthermore, we will stress the importance of quantifying uncertainty around causal effects and how to conduct sensitivity analyses of causal conclusions with regard to violations of key assumptions.

\section{The world of data science is about observational data}

Confounding bias is a key challenge when estimating causal effects from observational data. Let's assume that we are conducting an observational study to estimate the causal effect of a new drug compared to an older drug to lower blood pressure. Because the study is observational, it is highly likely that individuals that took the new drug are systematically different to the individuals that took the older drug with respect to their socioeconomic and health status. For example, it is possible that individuals with a higher income might have easier access to the new treatment and at the same time might be healthier than individuals with low income. Therefore, if we compare individuals taking the new drug to individuals taking the older drug without \emph{adjusting} for  income, we might conclude that the new drug is effective, when instead the difference we observe in blood pressure is due to individuals taking the new drug being richer and healthier to begin with.  

For now, we can assume that a variable is a potential confounder  if it is a pre-treatment characteristic of the subjects (e.g. income) that is associated with the treatment (e.g. getting the new drug) and also associated with the outcome (e.g. blood pressure) \footnote{A formal definition of a confounder is offered in Section 3.2, after the introduction of notation, postulates and assumptions.}. 

In our second example we define the treatment and the outcome as follows:
\begin{itemize}
\item Treatment - getting a dog $W=1$, not getting a dog $W=0$
\item Outcome - severe depression symptoms  $Y=1$, mild  depression symptoms  $Y=0$ measured after the treatment assignment $W$ 
\end{itemize}

Confounders could mask or {\it confound} the relation between $W$ and $Y$ which complicates causal attribution or leads to potentially incorrect inferences. For the depression/dog example (Figure 1), a potential confounder is the severity of depression symptoms (denoted by $X$) before treatment assignment. It is reasonable to believe that individuals with severe symptoms of depression pre-treatment ($X=1$) are more likely to adopt a dog $(W=1)$ than people with mild symptoms of depression ($X=0$). Furthermore, individuals with severe symptoms of depression before the treatment assignment $(X=1)$ are more likely to have severe symptoms of depression after the treatment assignment $Y$, than individuals with mild symptoms of depression ($X=0$). 

RCTs are the gold standard study design used to estimate causal effects. To assess the causal effect on survival of getting a new drug compared to a placebo, we could randomize half of the patients enrolled in our study. Half would receive the new drug ($W=1$), and the other half would receive a placebo ($W=0$). Randomization is particularly important to establish the efficacy and safety of drugs (new and existing) \citep{Collins2020}. This is because randomizing patients eliminates systematic differences between treated and untreated observations. In other words, randomization ensures that these two sets of observations are as similar as possible with respect to all potential confounders, regardless of whether we measure these potential confounders, and are identical on average. If the distribution over the measured and unmeasured confounders are the same in the two groups, then we can use the treated observations to infer what would have happened to the untreated observations.

Unfortunately, randomization is often not possible, either because there are ethical conflicts (such as exposure to environmental contaminants) or because it is challenging to implement. In the latter case, the most constraining factors are the time and monetary expense of data collection. Additional limitations of randomization include inclusion criteria that are too strict and cannot study large and representative populations \citep{athey2017econometrics}. Moreover, inclusion criteria usually focus on simplified interventions (e.g., randomization to a drug versus placebo) that do not mirror the complexity of  real-world decision making. While the credibility (\textit{internal validity}) and ability to advance scientific discovery of RCTs is well accepted \cite[e.g., 2019 Nobel Memorial Prize in Economic Sciences,][]{duflo2007using, banerjee2015miracle}, there are large classes of interventions and causal questions for which results that have a causal interpretation can only be gathered from observational data.

Fortunately, in this new era of data science, we have access to significant observational data. We can, for example, easily identify large and representative populations of cancer patients, and determine from medical records who received standard therapy or new therapy (or multiple concomitant therapies). Additionally, we can ascertain age, gender, behavioural variables, income, health status before treatment assignment and assess cancer recurrence and survival \citep{Arvold:2014}. However, because we {\it observe} who receives treatment instead of {\it randomizing} who receives treatment, the treated and untreated sub-populations are likely to be systematically different with respect to each of the potential confounders. Without adjusting for systematic differences between the treated and untreated populations, our inference on causal effects will be biased. 
Given the significant amounts of available data, it is tempting to use correlations observed in the data as evidence of causation; but strong correlation can lead to misleading conclusions when quantifying causal effects.

\begin{figure}
    \centering
    \includegraphics[width=15cm]{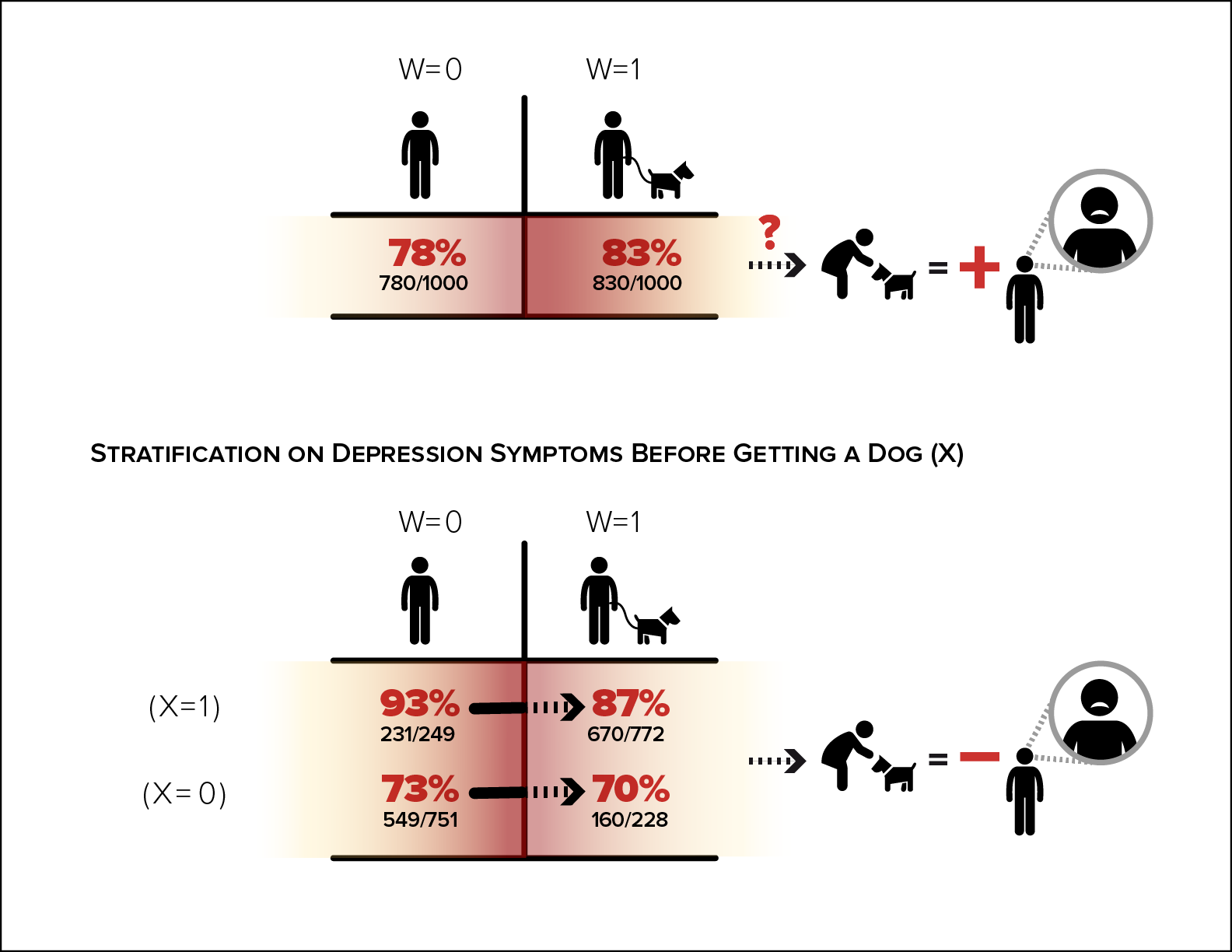}
    \caption{Top panel: rate of experiencing depression symptoms one year after getting a dog $W=1$ (83\%) and one year after not getting a dog $W=0$ (78\%). Bottom panel: rate of experiencing severe depression symptoms one year after getting a dog $W=1$ and one year after not getting the dog $W=0$, but separately for the two sub-populations that experienced severe or mild symptoms of depression before treatment assignment, denoted by $X=1$ and $X=0$, respectively. In the top panel, where we do not stratify by levels of $X$, getting a dog seems to increase the degree of severity of depression symptoms. In the bottom panel, where we stratify  by levels of $X$, getting a dog appears to decrease the severity of depression symptoms.}
    \label{fig:visual1}
\end{figure}

Figure 1 provides an example of confounding. Let's assume we compare two samples: one that adopts a dog $(W=1)$ and one that does not $(W=0)$. Then within each of these two populations, we calculate the rate of experiencing severe symptoms of depression $Y=1$. We found that adopting a dog appears to make the symptoms of depression worse: if you have a dog you are 5\% more likely ($83\%$ versus $78\%$) to experience severe symptoms of depression. Should we advise people not to own dogs? The problem with this analysis is that we ignore the fact that the subjects might be different in ways that would bias the conclusions. As mentioned before, a key potential confounder is the degree of severity of their depression symptoms {\it before} they were assigned the treatment ($X$). For example, let's stratify the two populations (treated and untreated) based on whether they experience severe or milder depression symptoms of ($X=1$ versus $X=0$) before treatment assignment. We find that within these two population strata, adopting a dog reduces the rate of experiencing severe symptoms of depression. 

This is an example of what is known as Simpson's paradox. Here the paradox occurs because people with severe depression symptoms before treatment assignment are more likely to adopt a dog. If we define by $e_i=P(W_i|X_i)$ the propensity of adopting a dog conditional to the level of depression symptoms pre-treatment assignment, then in this example $P(W_i=1|X_i=1)=772/(772+249)$ is higher than $P(W_i=1|X_i=0)=228/(228+751)$. In other words, the assignment to treatment, who gets a dog and who does not, is not completely random, as in a RCT. It is influenced by the pre-existing level of depression of the study subjects. Situations like these are very common in observational studies. We argue that the potential outcome (PO) framework detailed below allows us to design an observational study and clarify the assumptions that are required to estimate the causal effects in these studies. Such assumptions translate expert knowledge into identifying conditions that are hard, if not impossible, to verify from data alone. 

The PO framework is rooted in the statistical work on randomized experiments by \cite{Fisher1918, Fisher1925} and \cite{neyman1923}, extended by \cite{rubin1974estimating, Rubin1976,  rubin1977assignment, rubin1978bayesian, rubin1990formal} and subsequently by others to apply to observational studies. This perspective was called  {\it Rubin's Causal Model} by \cite{holland1986statistics} because it viewed causal inference as the result of missing data, and proposed explicit mathematical modeling of the assignment mechanism to reveal the observed data.

\section{The Design Phase of a Study}

We begin this section by describing the key distinctions between an RCT and an observational study. Table \ref{Tab: table1} summarizes and contrasts the main differences between RCTs and observational studies, and includes guidance on how to conduct causal inference in the context of observational studies in the last column.

The appeal of the RCT is that the design phase of the study (e.g. units, treatment, and timing of the assignment mechanism) is clearly defined {\it a priori} before data collection, including how to measure the outcome. In this sense, the RCT design is always prospective: treatment assignment randomization always occurs before the outcome is measured. A key feature of RCTs is that the probability of getting the treatment or the placebo, defined as the propensity score, is known -- under the experimenter's control -- and it does not depend on unobserved characteristics of the study subjects. Randomization of treatment assignment is also fundamentally useful because it balances observed and unobserved covariates between the treated and control subjects. Once the design phase is concluded, the experimenter can proceed with the analysis phase, that is, estimating the causal effects based on a statistical analysis protocol that was pre-selected without looking at the information on the outcome. The separation between design and analysis is critical as it guarantees {\it objective} causal inference. In other words, it will prevent the experimenter from picking and choosing the estimation method that would lead to their preferred conclusion \citep{rubin2008objective}.

In observational studies the treatment conditions and the timing of treatment assignment are observed after the data have been collected. The data are often collected for other purposes -- not explicitly for the study. As a result the researcher does not control the treatment assignment mechanism. Moreover, the lack of randomization means there is no guarantee that covariates are balanced between treatment groups which could result in systematic differences between the treatment group and the control group.
Traditionally, practitioners do not draw a clear distinction between the design and analysis phase when they analyze observational data. They estimate causal effects using regression models arbitrarily choosing covariates. This lacks the clear protocol of RCTs. To make objective causal inference from observational studies, we must address these challenges. Luckily, it is possible to achieve objective causal inferences from observational studies with careful design that approximates a hypothetical, randomized experiment. A carefully designed observational study can duplicate many appealing features of RCTs, and provide an objective inference on causal effects \citep{rubin2007design, rubin2008objective, hernan_robins_2016}. In the context of causal inference, the design of an observational study involves at least three steps (presented below). These steps should be followed by an analysis phase where the estimation approach is defined according to a pre-specified protocol as in the RCT. The three steps in the design phase are:
\begin{enumerate}
    \item define experimental conditions and potential outcomes (subsection 3.1);
    \item define causal effect of interest, including assumptions for identifiability  (subsection 3.2);
    \item construct a comparison group (subsection 3.3).
\end{enumerate}

\begin{table}\caption{Randomized Control Trials versus Observational Studies \label{Tab: table1}}
\begin{center}
\begin{tabular}{l | l | l || l}
\hline
& Randomized   & Observational  &    Best practices for causal inference in \\
Study characteristics	&   Control Trials &   Studies & observational studies\\
\hline
\textit{Basic concepts} &  &  \\
\quad Units  &     Defined before  &  Generally   & \multirow{2}{*}{Define units and treatments}\\
\quad Treatment &  data are collected & not specified &   \\
\quad Timing of treatment && &Determine timing of treatment   \\
\quad  assignment &&&assignment relative to measured \\\ &&& variables\\
&&\\
\textit{Design phase} &   &   The separation   & Hide outcome data \\
\quad \textit{versus} &   Separated  & is unclear  & until design phase\\
\textit{Analysis phase} &  & &  is complete \\
&&\\
\multicolumn{2}{l}{\textit{Treatment assignment mechanism}}&\\
\quad Unconfounded  & \multicolumn{1}{c|}{ \checkmark}& \multicolumn{1}{c||}{ ?} & Do data contain key confounders? \\
\quad Probabilistic& \multicolumn{1}{c|}{ \checkmark}& \multicolumn{1}{c||}{ ?} & \quad Yes: Assume unconfoundedness \\
\quad Known & \multicolumn{1}{c|}{ \checkmark}   &  \multicolumn{1}{c||}{ ?} & \quad No: Give up, \\ &&&  \quad find a better data source,\\ &&&  \quad or use a different identification strategy\\
&&\\
\textit{Covariate balancing} & Guaranteed   & Not guaranteed  & 
1. Assess overlap in covariates \\ 
&&& \, \,\, distribution \\
& (in expectation) &   & 2.  Remove units not similar to any\\ &&&\,\, \,   units in the opposite treatment group\\
&&& 3. Find subgroups in which the  \\ &&&\,\, \, treatment groups are balanced on \\ &&&\,\, \, covariates \\
&&\\
\textit{Type of analysis} &  By protocol & Regression  & Analyze according to \\ &&&a pre-specified protocol\\ &&& Complement results with sensitivity analysis \\ &&& to deviations from assumptions\\
\hline
\end{tabular}
\end{center}
\end{table}

\subsection{Define the experimental conditions and the potential outcomes}

The first step to address a causal question is to identify the conditions (actions, treatments, or exposure levels) required to assess causality. To define the causal effect of $W=1$ versus $W=0$ on $Y$, one must postulate the existence of two potential outcomes: $Y(W=1)$ and $Y(W=0)$. As the name implies, both variables are potentially observable, but only the variable associated with the observed (or assigned) action will be observed. The critical feature of the notion of a cause is that the value of $W$ for each unit can be manipulated. 
To illustrate this idea, we now introduce the subscripts $(t-1)$ and $t$ to indicate time at or before the treatment assignment and time after treatment assignment (e.g., one year later). For example, let's say Amy (subject $i$ at a specific time $t-1$) adopted a dog ($W_{i,t-1}=1$) and we observe her depression symptoms one year later $t$ ($Y_{i,t}(W_{i,t-1}=1)=Y_{i,t}(1)$). Now we must assume that $W$ can be manipulated, that is, we must be able to hypothesize a situation were Amy would not get a dog ($  W_{i,t-1}=0$). So, $Y_{i,t}(1)$ is observed and $Y_{i,t}(0)$ is unobserved.

\subsection{Define the causal effect of interest} We define a unit-level causal effect as the comparison between the potential outcome under treatment and the potential outcome under control. For example, the causal effect of $W$ on $Y$ for unit $i$ is defined as $Y_{i,t}(1)-Y_{i,t}(0)$. 

\begin{figure}
    \centering
    \includegraphics[width=10cm]{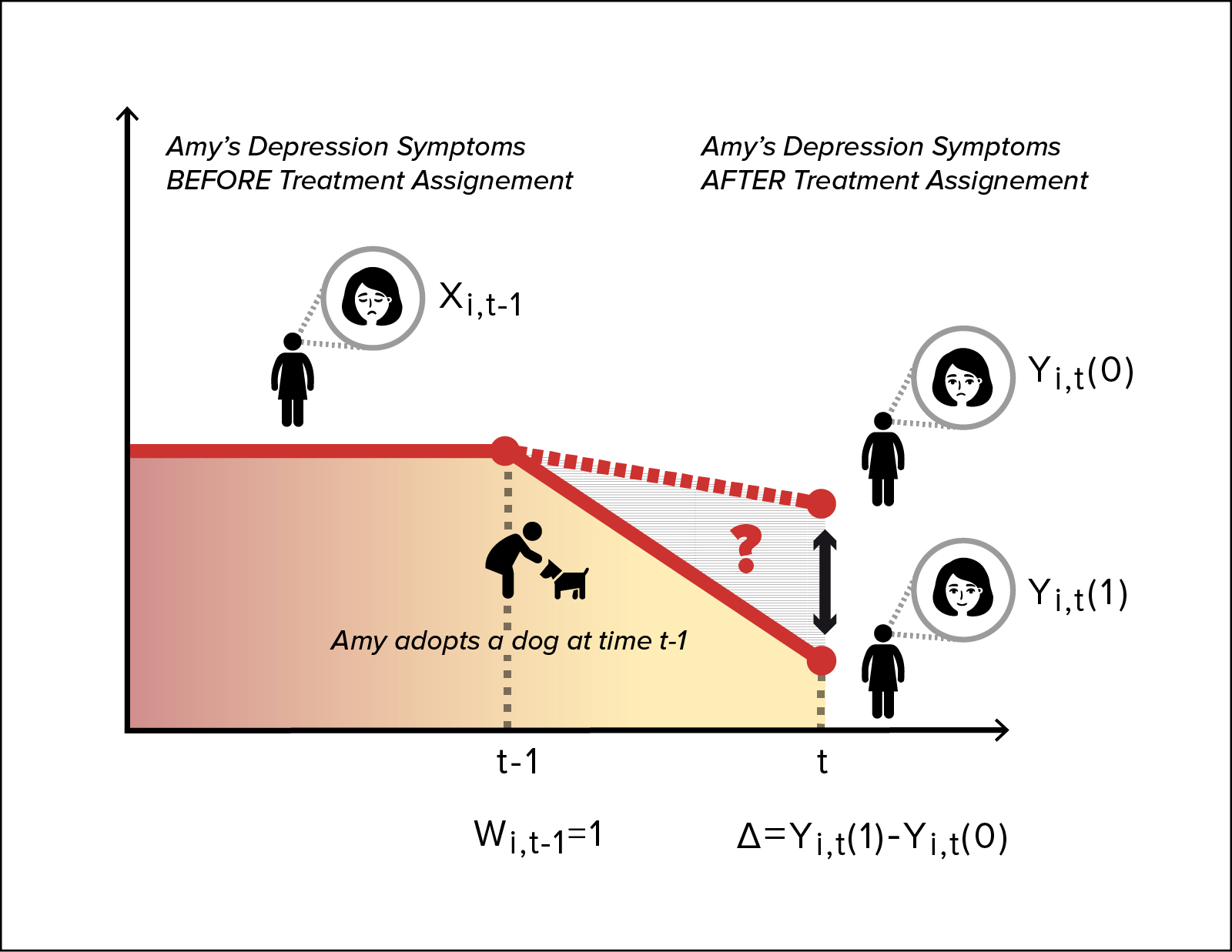}
    \caption{Causal effects as contrasts of potential outcomes at a given time point $t$.}
    \label{fig:visual2}
\end{figure} 

As we can see from Figure 2, the causal effect of interest, is not a pre/post comparison of the depression symptoms for Amy (defined as  $Y_{i,t}(1)-X_{i,t-1}$). It is, instead the difference between her two potential outcomes evaluated at time $t$, defined as  $Y_{i,t}(1)-Y_{i,t}(0)$, where $Y_{i,t}(1)$ is observed, whereas $Y_{i,t}(0)$ is not. These differences are called individual treatment effects (ITE) \citep{rubin2005}.  

While difficult to estimate \cite[e.g.,][]{Funk_et_al_2011}, ITEs inform treatment effect heterogeneity \cite[e.g.,][]{Issa_et_al_2016} and facilitate decision making in individualized settings where an estimate of the causal effect averaged across all the subjects in the sample may not be very practical \cite[e.g.,][]{li2019identify}.  We will return to this issue in Section 5. 
The last column of Table 2 introduces the concept of summarizing individual causal effects across the population of interest. For example, we can summarize the unit-level causal effects by taking the average difference, or by taking the difference in median values of the $Y_i(1)$s and $Y_i(0)$s, respectively. We typically focus on causal estimands that contrast potential outcomes on a common set of units (our target sample of size $N$), for example the average treatment effect (ATE): $\frac{1}{N} \sum_{i=1}^N(Y_i(1)-Y_i(0))= \bar{Y}(1)-\bar{Y}(0)$.

\begin{table}
\caption{The Science}
\begin{center}
{\normalsize
\begin{tabular}{c|c|c|c|c|c} \hline
Units & Covariates  & Potential Outcomes & Potential Outcomes & Unit-level& Summary of \\
 & $X$& $Y(1)$ & $Y(0)$ & Causal Effects & Causal Effects\\
 \hline
 1  & $X_1$ & $Y_1(1)$ & $Y_1(0)$ & $Y_1(1)$ vs $Y_1(0)$ & Comparison of \\
 . & . & . & . & . & $Y_1(1)$ vs $Y_1(0)$\\
$i$  & $X_i$ & $Y_i(1)$ & $Y_i(0)$ & $Y_i(1)$ vs $Y_i(0)$ & for a common \\
. & . & . & . & . & set of units\\
N  & $X_N$ & $Y_N(1)$ & $Y_N(0)$ & $Y_N(1)$ vs $Y_N(0)$ & \\
\hline \end{tabular}}
\end{center}
\end{table}

The fundamental challenge is that we will never observe a potential outcome under a condition
other than the one that actually occurred, so that we will never observe an individual causal effect (see Table 3). \cite{holland1986statistics} refers to this as the {\it fundamental problem of causal inference}. We typically refer to the missing potential outcome as the {\it counterfactual} and to the observed outcome as the {\it factual} outcome. Causal inference relies on the ability to predict the counterfactual outcome. It is important to note that methods used to predict or impute counterfactuals are different than off-the-shelf prediction or imputation often used for missing values. This is because we will never be able to find data where both potential outcomes $(Y_i(1), Y_i(0))$ are simultaneously observed on a common set of units. Table 3 highlights other fundamental implications of this representation of causal parameters: a) Uncertainty remains even if the $N$ units are our finite population of interest, because of the missingness in the potential outcomes; b) The inclusion of more units provides additional information (more factual outcomes) but also increases missingness (more counterfactual outcomes).

Data alone is not sufficient to predict the counterfactual outcome. We need to introduce several assumptions that essentially embed subject matter expert knowledge \citep{angrist_mostly_2008}. This is why machine learning alone cannot resolve causal inference problems, an issue discussed further in Section 5. To identify a causal effect from the observed data we have to make several assumptions. 

\begin{table}
\caption{What we are able to observe about the Science}
\begin{center}
{\normalsize
\begin{tabular}{c|c|c|c|c|c} \hline
Units & Covariates & Treatment&\multicolumn{2}{c|}{Potential Outcomes}& Unit-level \\
 & $X$& $W$ & $Y(1)$ & $Y(0)$ & Causal Effects \\
 \hline
 \vspace{-0.2cm}\\
 $1$  & $X_1$ & 1 & $Y_1(1)$ & ? & ?  \\
 \vdots &\vdots & \vdots & \vdots & \vdots & \vdots \\
$i$  & $X_i$ &0 & ? & $Y_i(0)$ & ?  \\
 \vdots & \vdots & \vdots & \vdots & \vdots & \vdots \\
$N$  & $X_N$ & 1 & $Y_N(1)$ & ? & ? \\
 \vspace{-0.2cm}\\
 \hline
\end{tabular}}
\end{center}
\end{table}

\noindent \textbf{Assumption 1: Stable Unit Treatment Value Assumption (SUTVA)}. SUTVA, introduced and formalized in a series of papers by Rubin \cite[see][]{rubin1980randomization, rubin1986comment, rubin1990formal}, requires that there is no interference and no hidden version of the treatment. No interference assumes that the potential outcomes of a unit $i$ only depend on the treatment unit $i$ receives, and are not affected by the treatment received by other units.
 
For example epidemiological studies of the causal effects of non-pharmaceutical interventions (e.g., stay-at-home advisory) on the probability of getting COVID19  violate the assumption of no interference. This is because the individual level outcome (whether or not a subject is infected) depends on whether he/she complies with the stay-at-home advisory, but also on whether or not others in the same household also comply with the stay-at-home advisory.

Spillover effects are a violation of the no interference assumption. These examples often occur when the observations are correlated in time or space. In our ice cream case study this assumption is likely to hold as it is reasonable to assume that the only person who benefits from weight loss is the person that stopped eating ice cream. SUTVA violations can be particularly challenging in air pollution regulation studies as pollution moves through space and presents a setting for interference. Intervening at one location (e.g., a pollution source) likely affects pollution and health across many locations, meaning that potential outcomes at a given location are probably functions of local interventions as well as interventions at other locations \citep{papadogeorgou2019causal,forastiere2020identification}. The condition of no hidden version of treatments requires that potential outcomes not be affected by {\it how} unit $i$ received treatment. This assumption is related to the notion of consistency \citep{hernan2016does, hernan_robins_2020}. For example, how Amy adopted a dog (a friend giving away puppies or driving to a breeder) does not affect Amy’s outcome.

Our ability to estimate the missing potential outcomes depends on the treatment assignment mechanism. That, is, it depends upon the probabilistic rule $W=1$ versus $W=0$ which determines whether $Y(1)$ or $Y(0)$ is observed. The assignment mechanism is defined as the probability of getting the treatment conditional on $X, Y(1), Y(0)$, e.g. $P(W| X, Y(1), Y(0))$. This expression will be simplified under the next assumption. 

\noindent \textbf{Assumption 2: No unmeasured confounding.} The assignment mechanism is unconfounded if: $P(W \mid X, Y(1), Y(0))=P(W \mid X)$. Unconfoundedness is also known as no unmeasured confounding assumption, or conditional independence assumption. This means that if we can stratify the populations within subgroups that have the same covariate values (e.g., same age, gender, race, income), then within each of these strata, the treatment assignment (e.g., who gets the drug and who does not) is  random.

The assumption allows to provide a formal definition of a confounder. Although there is no consensus regarding a unique and formal definition, we adopt the one proposed by \cite{vanderweele2013definition}: a pre-exposure covariate $X$ is said to be a confounder for the effect of $W$ on $Y$ if there exists a set of covariates $X^*$ such that the effect of $W$ on $Y$ is unconfounded conditional on $(X^*,X)$, but it is not for a subset of $(X^*,X)$. Equivalently, a confounder is a member of a minimally sufficient adjustment set.

Unconfoundedness is critical to estimate causal effects in observational studies. As $Y_i(1)$ is never observed on subjects with $W_i=0$ and $Y_i(0)$ is never observed on subjects with $W_i=1$, we cannot test this assumption, and so its plausibility relies on subject-matter knowledge. As a result, sensitivity analysis should be conducted routinely to assess how the conclusions will change under specific deviations from this assumption (discussed in Section 6). Moreover, this assumption may fail to hold if some relevant covariates are not observed, or if decisions are based on information on potential outcomes. For example a {\it perfect doctor} \citep{imbens2015causal} gives a drug to patients based on who benefits from the treatment (e.g., $W_i=I(Y_i(1)>Y_i(0))$): the assignment is confounded -- i.e. depends on the potential outcomes, irrespective of the covariates we are able to condition on. We discuss these situations in our final remarks. 

\noindent \textbf{Assumption 3: Overlap or positivity.} We define the propensity score for subject $i$ as the probability of getting the treatment given the covariates \citep{rosenbaum1983central} $e_i=P(W_i=1\mid X_i)$. The assumption of overlap requires that all units have a propensity score that is between 0 and 1, that is, they all have a positive chance of receiving one of the two levels of the treatment.

In the depression/dog example, this may be violated if some people in the population of interest are allergic to dogs and therefore their probability of getting a dog is zero. In the clinical example, this hypothesis is commonly violated if a patient has a genetic mutation that prevents him/her from receiving the treatment being tested. Because the propensity score can be estimated from data, we can check if overlap holds. If for some units the estimated $e_i$ is very close to either $1$ or $0$, then these units are only observed under a single experimental condition and therefore contain very little information about the causal effect. In this situation, strong assumptions are necessary. For example, a strong assumption is to say that the functional form that relates the covariates with the outcome holds also outside of the observed range of the covariates. A more formal approach of how to overcome violations of the positivity assumption is presented in \cite{nethery2019estimating}. If assumptions 2 and 3 are both met, then we conclude that the assignment mechanism is {\em strongly ignorable} \citep{rosenbaum1983central}. Classic randomized experiments are special cases of strongly ignorable assignment mechanisms. 

\subsection{How to construct an adequate comparison group} Once you have identified relevant potential confounders, and assuming they are sufficient for unconfoundedness to hold, the issue of confounding can be resolved by constructing an adequate comparison group. This is a crucial step in the design of an observational study. Our goal is to synthetically recreate a setting that is very similar to a randomized experiment, so the joint distribution of all the potential confounders is as similar as possible between the treatment and control groups \citep{ho2007matching, stuart_rubin_2008, stuart2010matching}. For instance, let's return to our example about Amy (in this case we drop the subscript $t$). Amy (subject $i$) got a dog $W_i=1$, so we observe $Y_{i,obs}=Y_i(W_i=1)$ whereas $Y_{i,mis}=Y_i(W_i=0)$. To estimate a causal effect we need to predict $Y_i(W_i=0)$, that is, what we would expect the severity of Amy's symptoms to be under the hypothetical (not observed) scenario in which Amy did not get the dog ($W_i=0$). To predict the missing counterfactual $Y_i(W_i=0)$, we need to define an adequate control group. Based on the considerations made so far, we need to find subjects that are similar to Amy with respect to the potential confounders (age, race, income, health status, severity of depression symptoms) before treatment assignment. The only difference between the matched subjects and Amy is that they did not get a dog. This should be done for Amy and for any subject in our target population who got a dog. With a large number of confounders matching on all confounders {\it exactly} may not be feasible. A common approach to address this challenge is to use propensity scores ($e_i$) and match subjects with respect to $e_i$. The estimated propensity score is an univariate summary of all covariates and is crucial to estimate causal effects under unconfoundedness \citep{Robins_et_al1995, RobinsRotnitzky1995, imbens2015causal, imai2014covariate_balance_PS}.  Subjects sharing the same value of the propensity score have the same distribution of the observed potential confounders whether they are treated or not. Estimated propensity scores can be applied in the design phase to assess overlap and construct a comparison group through matching, stratifying, or weighting observations \citep{rubin2008objective}.
Covariate balance can also be viewed as an optimization problem. Procedures based on this idea either directly optimize weights or find optimal subsets of controls such that the mean or other characteristics of the covariates is the same in the treatment and control group \citep{zubizarreta2012using, diamond2013genetic, zubizarreta2014matching, zubizarreta2015stable, li_balancing_2016, li2018addressing}.
 
To this point we have only discussed what we are interested in estimating, the study design, and the problem of confounding. Now we will discuss how we estimate the causal effects. Being able to identify causal effects is a feature of the causal reasoning used in the potential outcome framework. Without including causal reasoning in the design phase you cannot to recover the causal effect even with the most sophisticated machine learning or nonparametric methods \cite[e.g.,][]{MatteiMealli2015Discussion}.

\section{Estimation}
Causal estimands such as $ATE=\overline{Y}(1)-\overline{Y}(0)$ are a function of the observed $Y_{obs}$ and the missing potential outcomes (the counterfactuals) $Y_{mis}$. Therefore, an estimation strategy needs to implicitly or explicitly impute $Y_{mis}$. What follows is not a comprehensive review of all estimation methods of causal effects, but rather key ideas of Bayesian estimation of the average treatment effect \citep{rubin1978bayesian, imbens2015causal, dingli2018}.

\subsection{Bayesian methods for the imputation of the missing counterfactuals after the design phase is concluded}
Within the model-based Bayesian framework for causal inference \citep{rubin1975bayesian, rubin1978bayesian}, the $Y_{mis}$ are considered unknown parameters. The goal is to sample from their posterior predictive distribution conditionally to the observed data defined as: 
\begin{equation}
    P(Y_{mis}|Y_{obs},X,W)\propto P(X,Y(1),Y(0)) P(W|X,Y(1),Y(0))
\end{equation}
where $P(X,Y(1),Y(0))$ denotes the model for the potential outcomes, while $P(W|X,Y(1),Y(0))$ denotes the model for the treatment assignment. By sampling from this posterior distribution we can multiply impute $Y_{mis}$, and then estimate $ATE$, or any other causal contrast, and its posterior credible interval \citep{rubin1978bayesian,mealli2011modern}. Note that this missing data imputation exercise is critically different from a usual prediction task: the two models introduced above contain expert knowledge (for example the assumption of strong ignorability or the inclusion of relevant covariates) that cannot be retrieved from data alone. Under unconfoundedness or no unmeasured confounding (Assumption 2: $P(W|X,Y(1),Y(0))= P(W|X)$), and assuming the parameters of the model for the potential outcomes are a priori independent of the parameters of the model for the assignment mechanism, then the posterior distribution of the missing potential outcomes only depends on the parameters of the model for the outcomes. Specifically, assuming {\em exchangeability} \citep{Finetti63}, there exists a parameter vector $\theta$ having a known prior distribution $p(\theta)$ such that:
\begin{eqnarray}\label{definetti}
P(Y(0), Y(1), W, X) &=& \int \prod_i P(Y_i(0),Y_i(1),W_i,X_i,\theta)p(\theta)~d\theta.
\end{eqnarray}
The posterior predictive distribution of the missing data, $P(Y_{mis}\mid Y_{obs}, W, X)$, can be written as
\begin{eqnarray} \label{AM}
\lefteqn{P(Y_{mis}\mid Y_{obs}, W, X)} \\
&=&\frac{P(Y(0),Y(1), W, X)}{\int P(Y(0),Y(1), W, X) d Y_{mis}} \nonumber \\
&=&\frac{\int \prod_i P(W_i \mid Y_i(0),Y_i(1),X_i, \theta)
P(Y_i(0),Y_i(1)\mid X_i, \theta) P(X_i \mid \theta)p(\theta) d
\theta}{\int\!\!\int \prod_i P(W_i\mid
Y_i(0),Y_i(1), X_i,\theta) P(Y_i(0),Y_i(1)\mid
X_i,\theta)P(X_i\mid \theta)p(\theta)~ d \theta d Y_{mis}}.
\nonumber
\end{eqnarray}
Let $\theta_{W|X}$, $\theta_{Y|X}$ and $\theta_{X}$, denote the unknown parameters corresponding to the distribution of the treatment assignment mechanism (e.g. the propensity score), the distribution of potential outcomes, and the  distribution of covariates, respectively. Then, given ignorability, the propensity score $P(W_i\mid X_i,\theta_{W|X})$ and the covariates' distribution $P(X_i\mid\theta_{X})$ cancel out in Equation \eqref{AM}, which simplifies to:
\begin{eqnarray} \label{AM_reduced}
P(Y_{mis}\mid Y_{obs}, W, X)
&\propto& \int \prod_i P(Y_i(0),Y_i(1)\mid X_i, \theta_{Y|X})p(\theta_{Y|X}) d \theta_{Y|X} 
\end{eqnarray}
Equation \eqref{AM_reduced} shows that, under ignorable treatment assignments, the potential outcome model needs to be specified $P(Y_i(w)\mid X_i,\theta_{Y|X})$ for $w=0, 1$, as well as the prior distribution $p(\theta_{Y|X})$, to derive the posterior distribution of the causal effects.\footnote{Please refer to Section 7 in \cite{rubin1990formal} for a specific example, and \cite{ImbensRubin97} for the dependence of the posterior distribution of causal effects on association parameters in the joint distribution of $Y(1)$ and $(Y(0)$. The discussion is beyond the scope of this paper.} Therefore, the most straightforward Bayesian approach to estimate causal effects under ignorability is to specify  models for $Y(1)$ and $Y(0)$ that are conditional to covariates and some parameters and then draw the missing potential outcomes from their posterior predictive distribution, which will also derive the posterior distribution of any causal estimand.

As such, it seems like propensity scores, that are central to balance the covariates in the design stage, do not affect Bayesian inference for causal effects under ignorability.
However, as noted by \cite{rubin:1985}, for effective calibration of Bayesian inference of causal effects (i.e., good frequentist properties) good covariate balance is necessary \cite[see also]{ding2019bracketing}. That is, if covariates are balanced between treatment groups, inference on causal effects derived from the estimation of the outcome model are robust and not very sensitive to model assumptions. Flexible semi- and non-parametric specifications of the outcome model can be found in literature including Bayesian Additive Regression Models \cite[BART;][]{hill2011bayesian, hahn2020} or Bayesian Cubic Splines \citep{ChibGreenberg2010}.

\subsection{Bayesian methods for joint estimation of the outcome and propensity score models and the feedback problem} Some Bayesian methods explicitly include the estimation of the propensity score in the causal effects' estimation procedure. Some
of these approaches involve the specification of a regression model for the outcome with the propensity score as a single regressor, arguing that this modelling task is simpler than specifying a model for the outcome conditional on the whole set of (high-dimensional) covariates \citep{rubin:1985}. This approach can be improved by adjusting for the residual covariance between $X$ and $Y$ at each value of $e(X)$ \citep{gutman2015estimation}, or by specifying an outcome model conditional on  the covariates $X$ and the propensity score \citep{Zhen:Little:infe:2005}.

These are two-stage methods that separate design (estimation of the propensity score) from analysis. Some authors have proposed a single step Bayesian approach to merge the two stages. For example, \cite{McCandless09} proposed a joint modelling approach to estimate the outcomes and the propensity score in this setting. However \cite{zigler2013model} has shown that merging the two steps may induce the ``feedback problem" if the parameters of the two models are \textit{a priori} dependent. In this instance, the outcome information enters the estimation of the propensity score, contradicting ignorability, and can lead to a distorted estimate of the propensity scores and compromise estimates of causal effects. Consequently, these types of Bayesian models are not often used. \cite{liao2018uncertainty} considers how uncertainty associated with the design stage impacts estimation of causal effects in the analysis stage. Uncertainty in the design stage may stem from the propensity score estimation but also from how the propensity score is used. Liao and Zigler propose a procedure that obtains the posterior distribution of causal effects after marginalizing distributed over design-stage outputs. 

\subsection{Bayesian methods to overcome violations of the positivity} To address violations of the positivity assumption Bayesian methods, including BART and splines, have also been developed. For example, \cite{nethery2019estimating} have recently developed a novel Bayesian framework to estimate population average causal effects with minor model dependence and appropriately large uncertainties in the presence of non-overlap. In this approach, the tasks of estimating causal effects in the overlap and non-overlap regions are delegated to two distinct models, suited to the degree of data in each region. In this case, tree ensembles are used to non-parametrically estimate individual causal effects in the overlap region, where the data can speak for themselves. In the non-overlap region, where insufficient data support makes reliance on model specification necessary, individual causal effects are estimated by extrapolating trends from the overlap region via a spline model. The authors showed that their proposed approach has a good performance compared to approaches that perform estimation only in the area of overlap \citep{hill2013assessing,cefaluzigler2017}.

\subsection{Bayesian and non-Bayesian methods that are model-free} Estimation methods under ignorability that are model-free (i.e., matching, stratification, weighting) are usually frequentist and treat potential outcomes as fixed values instead of random variables \cite{imbens2015causal, imbens2004,yao2020survey}. Some authors have proposed quasi-Bayesian approaches that involve only the Bayesian estimation of the propensity score, and then estimate the causal effect via matching \citep{An10}, stratification or weighting \citep{Saarela2015}. These approaches are not fully Bayesian in that they incorporate only the uncertainty in the propensity score estimation and not in the imputation of the missing potential outcomes. The frequentist methods have been improved by combining them with outcome regression adjustments \cite[e.g.,][]{abadieimbens2011}. Robins and colleagues \cite[e.g.,][]{Scharfstein1999, Robins2000,lunceford2004stratification, bang2005doubly,RobinsRotnitzky1995,Robins_et_al1995, Funk_et_al_2011,knaus2020double} have proposed a class of double robust (DR) estimators that combine inverse probability weighting estimator (IPW) with an outcome regression. Interestingly, \cite{Gustafson12a} casted DR estimator from a Bayesian perspective as a weighted average of a parametric model and a satured model for the outcome conditional on covariates, with weights that depend on how well the parametric model fits the data. As a result, it can also be viewed as a Bayesian model average estimator \citep{cefalu2016model}.

\subsection{Bayesian and non-Bayesian methods to account for variable selection either in the propensity or in the outcome model} \cite{zigler_dominici_2014} proposed a Bayesian model averaging method to adjust for the uncertainty in the selection of the covariates in the propensity score model, extending \cite{wang2012bayesian}; see also \cite{wang_et_al_2015}. When the number of the potential confounders is larger than the number of observations, approaches for dimension reduction and penalization are required. The standard approaches (e.g., the Lasso) generally aim to predict the outcome, and are less suited for estimation of causal effects. Under standard penalization approaches, if a variable $X$ is strongly associated with the treatment $W$ but weakly associated with the outcome $Y$, its coefficient will be reduced towards zero leading to confounding bias. \cite{belloni2014inference, belloni2014high} proposed a modified version of Lasso, called double Lasso, to reduce confounding bias. There are several Bayesian alternatives that usually outperform such approaches, for example using continuous spike and slab priors on the covariates' coefficients in the outcome and propensity score models \citep{wang_et_al_2015, cefalu2016model,antonelli2017high, antonelli2018bayesian}.

\section{The perils and the strengths of machine learning methods in causal inference}
We have presented approaches for the estimation of $Y_{mis}$ with high dimensional covariates and nonparametric methods, and the related issue of variable selection. At first glance, these tasks could be addressed by implementing off-the-shelf machine learning methods, but there are challenges. In this section we provide critical insights regarding the application of machine learning methods in causal inference.

Machine learning methods primarily address prediction or classification problems, and  have been included in statistical textbooks \cite[e.g.,][]{hastie2009elements, James2013, efron2016computer}. On a high level, there are two broad categories of machine learning methods: supervised and unsupervised learning. In supervised learning, the predictors (i.e. covariates, features) $X$ and the outcome $Y$ are both observed. The goal is to estimate the conditional mean of an outcome $Y$ given a set of covariates or features $X$, to ultimately predict $Y$. These methods include decision trees \cite[e.g.][]{BreFriOlsSto84a}, random forests \citep{breiman2001random}, gradient boosting \citep{friedman2001greedy}, support vector machines \citep{cortes1995support, suykens1999least}, deep neural networks \cite[e.g.,][]{lecun2015deeplearning, farrell2018deep}, ensemble methods \cite[e.g.,][]{Dietterich00ensemblemethods}, and variable selection tools such as LASSO \citep{tibshirani1996regression, hastie2015statistical}.
Regression trees, and random forests as their extension, have become very popular methods for estimating regression functions in settings where out-of-sample predictive power is important. When the outcome $Y$ is an unordered discrete response, these supervised learning algorithms attempt to resolve classification problems - for example detecting spam emails. In this instance, a machine learning algorithm is trained with a set of spam-emails labelled as spam and not-spam emails labelled as not-spam, so that a new email can be classified as either spam or not-spam. In unsupervised learning \cite[see][]{hastie2009elements} only features $X$ are observed and the goal is to group observations into clusters \citep{jain1999data}. Clustering algorithms essentially group units based on their mathematical similarities and dissimilarities of features $X$. These tools can be used for example to find groups of basketball or soccer players with similar attributes and then interpret and use these clusters to form teams or to target coaching. Deep learning methods are another general and flexible approach to estimate regression functions. They perform very well in settings with extremely large number of features, like image recognition or image diagnostics \citep{he2016deep, simonyan2014very}. These methods typically require a large amount of tuning to work well in practice, relative to other methods such as random forests, and as a result we will not discuss them further in this article. 

Since machine learning supervised learning methods aim to estimate the conditional mean of an outcome $Y$ given $X$ it would on the surface appear to be a good fit to exploit machine learning methods to estimate the missing potential outcomes and as a result the causal effects, especially in high-dimensional settings. However, it is not that simple. The following section presents the circumstances under which off-the-shelf supervised learning machine learning methods might not be appropriate with regard to causal inference. We also discuss how these methods can be adapted to achieve estimation of causal effects; how machine learning and the literature on statistical causal inference can cross-fertilize; and the open questions and problems that machine learning cannot handle on its own.

\subsection{Why off-the shelf machine learning techniques might not be appropriate for causal inference.} A key distinction between causal inference and machine learning is that the former focuses on estimation of the missing potential outcomes, average treatment effects, and other causal estimands, and machine learning focuses on prediction and classification. Therefore machine learning dismisses covariates with limited prediction importance. However, in causal inference if these same covariates are correlated with the treatment assignment they can be important confounders. As previously discussed omitting covariates from the analysis that are highly correlated with the treatment can introduce substantial bias in the estimation of the causal effects even if their predicting power is weak. Another major difference is that machine learning methods are typically assessed on their out-of-sample predictive power. This approach has two major drawbacks in the context of causal inference. First, as pointed out by \cite{athey2016}, a fundamental difference between a prediction and the estimation of a causal effect is that in causal inference we can never observe the ground truth (e.g., in this context the counterfactual). That is, in our example, because Amy has adopted a dog, we will never be able to measure the severity of the Amy's depression symptoms under the alternative hypothetical scenario where Amy did not adopt a dog. Therefore, standard approaches for quantifying the performance of machine learning algorithms cannot be implemented to assess the quality of prediction of the missing potential outcomes and therefore the causal effects \citep{Gao2020}. Second, in causal inference we must provide valid confidence intervals for the causal estimands of interest. This is required to make decisions regarding which treatment or treatment regime is best for a given unit or subset of units, and whether a treatment is worth implementing. Another limitation of machine learning techniques in causal inference is that they are developed for the most part in settings where the observations are independent and therefore have limited ability to handle data that is correlated in time and/or in space, such as time series, panel data, spatially correlated processes. Additionally, they are not able to handle specific structural restrictions suggested by subject-matter knowledge, such as monotonicity, exclusion restrictions, and endogeneity of some variables \citep{angrist_mostly_2008, athey2019machine}.  

\subsection{How machine learning methods can adapt to the goal of estimation of causal effects.}

{Machine learning methods have been adapted to address causal inference. One popular approach is to redefine the optimization criteria, which typically depend on a function of the prediction errors, to prioritize issues arising in causal inference, such as the controlling for confounders and the discovering of treatment effect heterogeneity \citep{Chernozhukov2017, chernozhukov2018double}. For example, the causal tree method proposed by \cite{athey2016} is based on a rework of the criterion function of Classification and Regression Trees \citep{BreFriOlsSto84a} -- originally aimed at minimising the predictive error -- to maximise the variation in the treatment effects and, in turn, discover the subgroups with the highest heterogeneity in the causal effects (further details are presented in  Section 6). 
Typical regularizing algorithms used in machine learning, such as Lasso, Elastic Net and Ridge \citep{hastie2015statistical}, must prioritize confounding adjustment to avoid missing relevant covariates, as seen in \cite{belloni2014inference} and \cite{belloni2014high} and discussed in Section 4.5. Additional research efforts are required to study the statistical properties of machine learning techniques, as in \cite{wager2018estimation} and \cite{athey2019generalized} for random forests and \cite{jeong2020art} for BART. Treatment effect estimators based on causal random forests have been shown to be point-wise consistent and asymptotically normal, while estimators based on BART are asymptotically optimal. Both these asymptotic results hold true under a set of specific conditions that are not necessarily mild. However, research has shown that machine learning algorithms show promise when the hardest issues are resolved during design \citep{dorie2019automated, hahn2019atlantic, bargagli2019heterogeneous, knaus2020machine, bargagli2020causal}. Specifically, it is critical to first identify a good control population, for example using propensity score matching as discussed earlier. After we have achieved this critically important task, then we can be confident that the assumptions of identifiability of causal effects are met. Only at this stage, the machine learning techniques provides an excellent tool to predict the missing counterfactuals  \cite[e.g.,][]{chernozhukov2018double}. However, it is important to keep in mind that performance will ultimately rely on the flexible parametrization that machine learning methods impose on the data, the plausibility of the unconfoundedness assumption, and the extent of the overlap in the distribution of the covariates \citep{hernan_hsu_healy_2019}. Even in the presence of a high-dimensional set of covariates, the study design is important \citep{damour2017overlap}. 

\subsection{Cross-fertilization between machine learning and causal inference problems}
Several machine learning techniques have been adapted to improve traditional causal inference methods. These include approaches to regularization (the process of adding information to solve ill-posed problems or to prevent overfitting) that scale well to large datasets \citep{chang2018scalable} and the related concept of sparsity (the idea that some variables may be dropped from the analysis without affecting the performance of estimation of causal effects). The use of model averaging and ensemble methods common in machine learning is a practice that is now exploited in causal inference (see Section 4) \citep{van2011targeted, cefalu2016model}.

Framing data analysis as an optimization problem has inspired the development of causal inference methods based on direct covariate balancing. For example, \cite{zubizarreta2015stable} which optimizes weights for each observation instead of trying to estimate the propensity score so that the covariate distribution is the same in the treatment and control group. 
Matrix completion methods were originally developed in machine learning for the imputation of the missing entries of a partially observed matrix \citep{candes2009exact, recht2011simpler}. These methodologies can be used to improve causal inference methods for panel data and synthetic control methods \citep{abadie2010synthetic} in settings with large $N$ and $T$. In particular, matrix completion can be successfully adapted for the imputation of missing counterfactuals when a large proportion of potential outcomes is missing \citep{athey2018matrix}.

\subsection{Open questions and problems that machine learning alone cannot handle.}

However, even when adapted to treatment effect estimation, machine learning algorithms must be implemented with extreme caution when there are unresolved key issues surrounding the study design. For example:
\begin{itemize}
    \item The sample under study is not representative of the population about which we need or want to draw conclusions; 
\item The number of potential confounders that are measured may not be sufficient: there is nothing in the data that tells us that unconfoundedness holds; causal effect estimation should be followed by a well-designed sensitivity analysis; 
\item The presence of  post-treatment variables that must be excluded (i.e., variables that can be affected by the treatment and are strong predictor of the outcome);
\item  The lack of overlap \citep{nethery2019estimating, damour2017overlap} in the distribution of the estimated propensity scores for the treated and untreated, demands the machine extrapolate beyond what is observed;
\item If interference is detected, causal estimands (direct and indirect - spillover - effects) need to be re-defined and different estimation strategies need to be implemented. \citep{arpino2016, forastiere2016identification, papadogeorgou2019causal, bargagli2020heterogeneous, tortu2020modelling}. 
\end{itemize}
Even in settings where machine learning can accurately predict potential outcomes \citep{wager2018estimation, hahn2020, belloni2014high, Chernozhukov2017, chernozhukov2018double}, they cannot handle other problems such as missing outcomes \citep{Hill2004ReducingBI, matteimeallipacini2014}, (non-random) measurement error or misclassification of outcome or treatment \citep{Imai2010, BrandonVanderWeele2012, FunkLandi2014}, censoring due to death of the outcome \citep{rubin2006, MatteiMealli:2007}, and disentangling different causal  mechanisms \citep{MatteiMealli:2011, vanderweele2015explanation, BacciniMatteiMealli2017, ForastiereMatteiDing:2018}.

\section{Treatment effect heterogeneity: is the treatment beneficial to everyone? }
Suppose we found statistically significant evidence that a new drug prolonged life expectancy on average for the population under study. Should we encourage anyone, regardless of their age, income, or other diseases to take this new drug? It is often highly desirable to characterize which subgroups of the population would benefit the most or the least from a treatment. These types of questions require us to analyze treatment effect heterogeneity based on pre-treatment variables. There is extensive literature on assessing heterogeneity of causal effects that is based on estimating the conditional average treatment effect (CATE), which is defined as $E[Y(W=1) \mid X=x]- E(Y(W=0) \mid X=x]$ where $Y(W=1)\mid X=x$ and $Y(W=0)\mid X=x$ are the potential outcomes in the subgroups of the population defined by $X=x$. Conditionally to $X=x$, the CATE can be estimated with the same set of the causal assumptions that are needed for estimating the ATE \citep{athey2016}. Recently, machine learning methods such as random forests, Bayesian Additive Regression Tree (BART) \citep{chipman2010bart}, and forest based algorithms \cite{foster2011subgroup} and \cite{hill2011bayesian}, have been used to estimate CATE, especially in the presence of high dimensional $X$. Despite accurately estimating the CATE using machine learning methods, these methods offer little guidance about which population subgroups are important in the treatment effect heterogeneity. Their parametrization of the covariate space is complicated and difficult to interpret even by human experts. We define interpretability as the degree to which a human can understand the cause of a decision or consistently predict the results of the model \citep{miller2018explanation, kim2016examples}. Decision rules fit well with this non-mathematical definition of interpretability. A decision rule consists of a set of conditions about the covariates, that define a subset of the features space, and correspond to a specific subgroup. In our recent work \citep{lee2020causal}, we propose a novel causal rule ensemble (CRE) method that ensures interpretability of the causal rules while maintaining a high level of accuracy of the estimated treatment effects for each rule. We argue that in the context of treatment effect heterogeneity, we want to achieve at least two main goals: to (1) discover de novo the rules (that is, interpretable population subgroups) that lead to heterogeneity of causal effects, and (2) make valid inference about the CATE with respect to the newly discovered rules. \cite{athey2016} has introduced a clever approach to make valid inferences in this context. They introduced the idea of a sample-splitting approach that divides the total sample into two smaller samples: (1) to discover a set of interpretable decision rules that could lead to treatment effect heterogeneity (i.e. discovery sample), and (2) to estimate the rule-specific treatment effects and associated statistical uncertainty (i.e. inference sample). This is a very active area of research and one where the integration of machine learning and causal inference could provide important advances.

\section{Sensitivity analysis}
Sensitivity analyses can be conducted to bound the magnitude of the causal effects as a function of the degree to which the assumptions are violated to evaluate the robustness of causal conclusions to violations of unverifiable assumptions \citep{imbens2015causal, ding2016sensitivity}. 
Sensitivity analysis is different from model assessment or model diagnostic, because identifying assumptions, such as unconfoundedness, are intrinsically untestable: the observed data are uninformative about the distribution of $Y(0)$ for treated units $(W=1)$ and about the distribution of $Y(1)$ for control units $(W=0)$. We like to argue that the larger the set of measured covariates the smaller the chance of unmeasured confounding bias, and it is thus often sensible to assume unconfoundedness. However, in practice, even if we have adjusted for all measured covariates, we can never rule out the possibility of an unmeasured confounder. A sensitivity analysis assumes the existence of at least one unmeasured confounder and posits assumptions on how it may relate to both treatment assignment and outcome. It then examines how the estimated causal effect varies as the degree of confounding bias, resulting from this unmeasured covariate, increases. Different strategies for sensitivity analysis have been proposed in the literature and include \cite{rosenbaum1987sensitivity, rosenbaum2002observational, imbens2003sensitivity, ichino2008temporary, Gustafson2010sensitivity, Liu2013sensitivity, ding2016sensitivity, VanderWeele2017SensitivityAI, Franks2019sensitivity, zhang2019semiparametric}.  Software has been developed to perform sensitivity analyses that assesses the strength of conclusions to violations of unverifiable assumptions \citep{Gang2004sensitivity, Nannicini2007, keele2014sensitivity, twang2017}. 

One type of sensitivity analysis is so-called {\em placebo} or {\em negative control} \citep{imbens2015causal}. Here the goal is to identify a variable that cannot be affected by the treatment. This variable is then used as an outcome (say $Y^\star$) and the causal effect of $W$ on $Y^\star$ is estimated, when we know that the true causal effect, say $\Delta$, should be zero. If we estimate that $\Delta$ is statistically significantly different from zero, when we have adjusted for all the measured confounders, then we can conclude that there is unmeasured confounding bias. For additional details see \cite{schuemie2020confident}.

Sometimes we can check the quality of an observational control group by exploiting useful comparisons. For example, we can have access to two different pools of controls and use them both to check robustness of results to the use of either one \citep{Rosenbaum:1987}. We can also assess the plausibility of unconfoundedness and the performance of methods for causal effects in observational studies by checking the extent to which we can reproduce experimental findings using observational data. For example, \cite{dehejia2002propensity} use data from \cite{lalonde:1986} evaluation of non-experimental methods that combine the treated units from a randomized evaluation of the National Supported Work (NSW) Demonstration, a labor training program, with non-experimental comparison units drawn from survey datasets. They show that estimates obtained with propensity score based estimators of the treatment effect are closer to the experimental benchmark estimate than other methods.

In addition to the challenge of adjusting for confounding bias, there is another key challenge that is often overlooked, and that is due to over adjustment.  For example, controlling for post-treatment variables (such as mediators) will bias the estimation of the causal effects see for example \cite{greenland1999causal,hernan2004structural,rubin2008objective, shrier2008propensity, schisterman2009overadjustment,ding2015adjust}.
In addition, it is common to include as many covariates as possible to control for confounding bias; however M-bias is a problem that may arise  when the correct estimation of treatment effects requires that certain variables are not adjusted for (i.e., are simply neglected from inclusion in the model). The topic of over adjustment is also broadly discussed in the causal graphical models literature \cite[see][]{Pearl1995, shpitser2010validity, vanderweele2011new, perkovic2017complete}. 

\section{Discussion}
This article has focused on the potential outcome framework as one of the many approaches to define and estimate the causal effects of a given intervention on a outcome. We have presented:
\begin{itemize}
\item Thoughts regarding the central role of the study design when estimating causal effects (Table 1). 
\item Why machine learning is not a substitute for a thoughtful study design, and that it cannot overcome data quality, missing confounders, interference, or extrapolation. 
\item That machine learning algorithms can be very useful to estimate missing potential outcomes after issues related to study design have been resolved. 
\item Machine learning algorithms show great promises in discovering de novo subpopulations with heterogeneous causal effects.
\item The importance of sensitivity analysis to assess how the conclusions are affected by deviations from identifying assumptions.
\end{itemize}

This review is focused on data science methods for \emph{prospective} causal questions, that is, on assessing the causal effect of a particular, sometimes hypothetical, manipulation. 

This is a different goal than that of causal discovery which investigates the causal structure in the data, without starting with a specified causal model. To read more about causal discovery we refer to \cite{Glymour:Scheines:Spirtes:2014,Mooij:Peters:Janzing:Zscheischler:Scholkopf:2016,SpirtesZhang2016}. 

We briefly outlined the Bayesian approach as one of the many statistical methods to estimate missing potential outcomes and the ATE. Alternative approaches such as those proposed by Fisher and Neyman are based on the randomization distribution of statistics induced by a classical randomized assignment mechanism \citep{neyman1923, fisher1937design} and their sampling distribution. The key feature of these approaches is that potential outcomes are treated as fixed but unknown, while the vector of treatment assignments, $W$, and the sampling indicators are the random variables. The concepts created by these methods, p-values, significance levels, unbiased estimation, confidence coverage, remain fundamental today \citep{Rubin2010Shadish}. However, we believe that the Bayesian thinking provides a straightforward approach to summarize the current state of knowledge in complex situations, to make informed decisions about which interventions look most promising for future application and to properly quantify uncertainty around such decisions.

There are settings or instances where adjusting for measured covariates is not enough, that is, we cannot rule out dependence of the assignment mechanisms on the potential outcomes. In these \textit{irregular} settings, another important area of research is relying on identification strategies that differ from strong ignorability, some of which are called quasi-experimental approaches. A natural experiment can be thought of as an observational study where treatment assignment, though not randomized, seems to resemble random assignment in that it is haphazard, not confounded by the typical attributes that determine treatment assignment in a particular empirical field \citep{zubizarreta2014isolation}. An example is instrumental variable (IV) methods \citep{angrist1996identification} where a variable, the instrument, plays the role of a randomly assigned incentive of treatment receipt and can be used to estimate treatment effect with a non-ignorable treatment assignment. For example, suppose we want to study the causal effect of an additional child on female labour supply; fertility decisions are typically \textit{endogenous} and plausibly determined by observed and unobserved characteristics. \cite{AngristEvans:1998} used the sex of the first two kids as an instrument for the decision to have a third child, and estimated the effect of having an additional child on a women's employment status. Such strategies are very popular in socioeconomic applications. In addition, other examples include regression discontinuity designs \citep{imbens2008regression, li2015}, synthetic controls \citep{abadie2010synthetic}, and their combinations \citep{athey2018matrix, arkhangelsky2019synthetic}. These designs typically focus on narrow effects (e.g., of compliers, of units at the threshold) with high internal validity, that need to be extrapolated to people or the population we are interested in. These methods can also be improved using machine learning ideas (for making IV stronger, or using more lagged values in a nonparametric way) but they require subject-matter knowledge and a level of creativity that, at least now, machine learning does not have.
 
 Another area of active research is how to extend or generalize results from an RCT to a larger population \citep{Stuart2018generalizability}. \cite{Hartman2015} spells out the assumptions for this generalization \citep{Pearl:2011wx} and proposes an approach to estimate effects for the larger population.

There are other key areas of causal inference not covered in this article. These include mediation analysis \citep{vanderweele2015explanation, huber2020mediation} and principal stratification \citep{frangakis2002principal, Mealli2012ARA}, both which provide understanding of causal mechanisms and causal pathways. We have also contributed to the literature in these areas \cite[e.g.,][]{MealliPacini:2013,mattei2013aoas, forastiere2016identification, MealliPaciniStanghellini:2016, BacciniMatteiMealli2017,mattei2020switching}. These methods attempt to address questions such as: why having a dog reduces the level of severity of the symptoms of depression? It is because they make you happier, or when you have a dog you are outside and exercising more which helps depression, or that dogs help boost our immune systems? These questions are about exploring treatment effect heterogeneity with respect to post-treatment variables (e.g., spending more time outdoors as a consequence of walking the dog is a post-treatment variable). Questions regarding mediation require different tools than regression, matching, and machine learning prediction. In this new era of data science, where we have access to a deluge of data on pre-treatment variables, complex treatments, post-treatment variables and outcomes, this area of causality will become more prominent.
 
Can we envision a future where all these steps, characterizing the design and the analysis of observational data, and the unavoidable subject-matter knowledge, be translated into meta data and automatized? Perhaps. But this a challenge and there is still a lot of fascinating work ahead of us.

\newpage
{\bf Acknowledgements:}

We sincerely thank Dr. Kevin Josey, Dr. Alessandra Mattei and Dr. Xiao Wu and Lena Goodwin for providing feedback and editing the manuscript. Funding was provided by the Health Effects Institute (HEI) grant 4953-RFA14-3/16-4, National Institute of Health (NIH) grants R01,
 R01ES026217,
 R01MD012769,
 R01 ES028033, 
 1R01AG060232-01A1, 
 1R01ES030616,
 1R01AG066793, 
 R01ES029950, the 2020 Starr Friedman Award, Sloan Foundation, and
EPA 83587201-0 }

%%%%%%%%%%%%%%%%%%%%%%%%%%%%%%%%%%%%%%%%%%%
%%%%%%%%%%%%%%%%%%%%%%%%%%%%%%%%%%%%%%%%%%%
%%%%%%%%%%%%%%%%%%%%%%%%%%%%%%%%%%%%%%%%%%%
%\pagebreak
\bibliographystyle{apalike}
\bibliography{references}
\pagebreak
%%%%%%%%%%%%%%%%%%%%%%%%%%%%%%%%%%%%%%%%%%%
%%%%%%%%%%%%%%%%%%%%%%%%%%%%%%%%%%%%%%%%%%%
%%%%%%%%%%%%%%%%%%%%%%%%%%%%%%%%%%%%%%%%%%
\end{document}